# Project-X: A New High Intensity Proton Accelerator Complex at Fermilab

R. Tschirhart
*Fermi National Accelerator Laboratory, Batavia, IL 60540, USA*

Fermilab has been working with the international particle physics and nuclear physics communities to explore and develop research programs possible with a new high intensity proton source known as ``Project-X''.  Project X will provide multi-megawatt proton beams from the Fermilab Main Injector over the energy range 60-120 GeV  simultaneous with multi-megawatt protons beams at 3 GeV with very flexible beam-timing characteristics.   The Project-X research program includes world leading sensitivity in long-baseline neutrino experiments, neutrino scattering experiments, a rich program of ultra-rare muon and kaon decays, opportunities for next-generation electric dipole moment experiments and other nuclear/particle physics probes that reach far beyond the Standard Model.

## 1.  Introduction

A recent review panel [1] of the High Energy Physics Advisory Panel within the US identified three frontiers of scientific opportunity for the field of particle physics: the energy frontier, the intensity frontier and the cosmic frontier. "Project X", a proposed new high-intensity proton source at Fermilab [2], has the potential to be the flagship of discovery at the intensity frontier. Project X would deliver very high-power proton beams at energies ranging from 3.0 to 120 GeV. It would also offer unprecedented flexibility in the timing structure of beams (pulsed or continuous wave, varying gaps between pulses, fast or slow spill) and in a variety of simultaneously delivered secondary beams. These features would make Project X the foundation both for fundamentally new experiments and for significant advances in ongoing experimental programs in neutrino physics and the physics of ultra-rare processes.

Physics at the intensity frontier is closely linked with both the energy and the cosmic frontiers. Answers to the most challenging questions about the fundamental physics of the universe will come from combining what we learn from the most powerful and insightful observations at each of the three frontiers. Addressing most of the questions under investigation at the energy and cosmic frontiers also requires measurements at the intensity frontier.

Understanding neutrinos and their masses, for example, may address the central question of the ultimate unification of forces. Matter-antimatter asymmetry in the behavior of neutrinos might elucidate one of the deepest mysteries of physics: why do we live in a universe made only of matter, with no antimatter? Results from experiments now underway around the world will shape the future course of neutrino research. No matter what they find, Project X, with the world's most intense neutrino beams, will be key to the next steps in neutrino physics.

Characterizing the properties and interactions of new particles that will likely emerge from discoveries at the Large Hadron Collider (LHC) will require the perspective of experiments at the intensity frontier. If experiments at the LHC discover super-symmetry, for example, intensity-frontier searches have the potential to make critical distinctions among different models of this phenomenon. And if LHC experiments should fail to see new physics, the intensity frontier would be the only approach to access new physics that may exist beyond the TeV scale.

High-intensity particle beams spur experimental investigations wherever dramatic advances require extraordinary precision and clean, background-free experimental conditions. They provide the capability for neutrino studies, such as long-baseline experiments, that require detailed, precise studies of energy spectra in order to detect matter-antimatter asymmetry in neutrinos. These beams allow physicists to focus sharply on barely observable processes with great scientific significance, such as the high-priority search for the coherent conversion of muons to electrons. In experiments now limited by statistics, such as the search for the transition of a quark of one flavor to an identically charged quark of a different flavor, (flavor changing neutral currents) very high intensity beams make possible the precise measurements that are essential for discovery. They provide the exacting experimental conditions required by extremely challenging experiments, such as the search for the rare decays of kaons which require high duty-factor beam trains with very fast (<50 psec) pulses within the train.  Precision measurement of the $K \rightarrow \pi \nu \bar{\nu}$ kaon decays are particularly promising probes of physics beyond well beyond the LHC's reach.  Finally, these beams make possible experiments that may be crucial for a true understanding of physical phenomena such as electric dipole moments of atoms, a very incisive probe of matter-antimatter asymmetry at energy scales beyond the Standard Model.

The concept of the Project X accelerator complex is illustrated in Figure 1 opens the window on a whole spectrum of new experiments at the intensity frontier. The scientific opportunities provided by Project X are in  four areas: neutrinos, muons, kaons, and fundamental physics using nuclear physics techniques. With the power of Project X, they all attain new, hitherto unattainable, capabilities for discovery. Project X would also represent a first step toward potential future particle physics facilities, such as a neutrino factory or an energy-frontier muon collider.

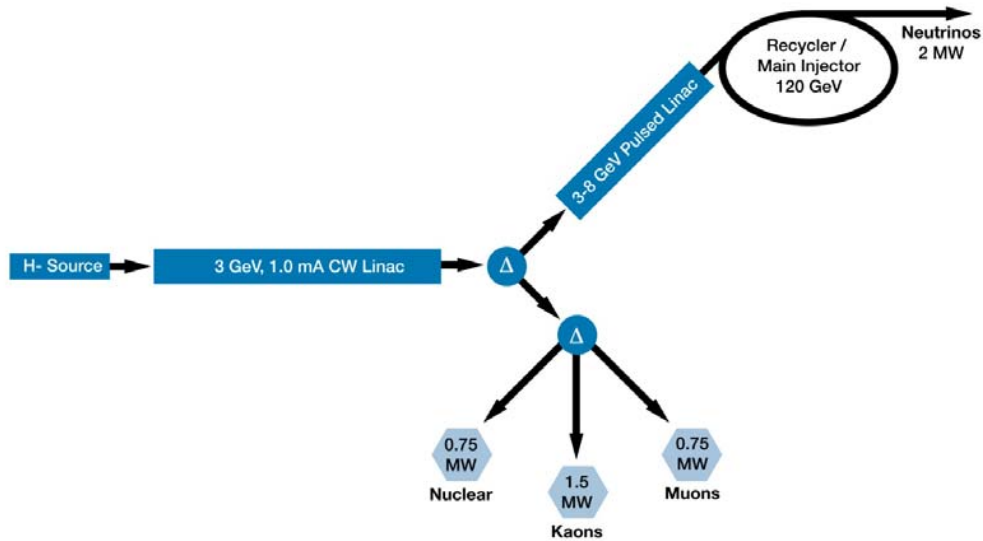

**Figure-1:** Project X, a high-power proton facility, would support world-leading programs in long-baseline neutrino physics and the physics of rare processes  Project X is based on a 3 GeV continuous-wave superconducting  H- linac. Further acceleration to 8 GeV, injected into Fermilab's existing Recycler/Main Injector complex, would support long-baseline neutrino experiments. Project X would provide 3.0 MW of total beam power to the 3 GeV program, simultaneously with 2 MW to a neutrino production target at 60-120 GeV and 200 kW at 8 GeV.

## 2.  Project-X and the Fermilab Accelerator Complex

For many years Fermilab has operated both the highest-energy particle collider and the highest-intensity accelerator based neutrino beam in the world. Now the LHC is surpassing the Tevatron in energy and Japan's J-PARC facility is embarking on a long-baseline neutrino program in strong competition with the Fermilab program. In this international context, the US elementary particle physics community has adopted a strategic plan for the coming decade that emphasizes research on three frontiers: the energy frontier, the intensity frontier and the cosmic frontier. The plan recognizes that over the coming decade Fermilab will be the sole US site for accelerator-based particle physics research. Fermilab's strategy is fully aligned with the US plan. It features the development of a high-intensity proton source as the key to the long-term US program.

Project X is a multi-MW proton accelerator facility proposed for construction at Fermilab.  It is based on an H- linear accelerator using superconducting RF technology. Project X would be the linchpin for future development of the Fermilab accelerator complex, providing long-term opportunities at both the intensity and energy frontiers. Project X would provide great flexibility for intensity-frontier physics, creating the opportunity for a long-term world leading program in neutrino physics and other beyond-the-standard-model phenomena.  The technology [3] for Project X also opens opportunities beyond traditional particle physics applications, including:

- Accelerator-driven energy systems

- Rare isotope production for nuclear physics

- Neutron sources

- X-ray FELs

- Energy recovery linacs

- Muon facilities for materials research

The technology development for Project X is also closely aligned with the technologies required for the proposed International Linear Collider, preserving Fermilab's capability to serve as a host, or major contributor, to such a possible future accelerator. The development of multi-MW capabilities could also provide the basis for a future muon collider.

## Summary

A "national collaboration with international partners" has formed within the US to develop Project X. The national collaboration comprises Argonne National Laboratory, Brookhaven National Laboratory, Cornell University, Fermilab, Lawrence Berkeley National Laboratory, Michigan State University, Oak Ridge National Laboratory, Thomas Jefferson National Accelerator Facility, SLAC National Accelerator Laboratory, and the Americas Regional Team of the ILC Global Design Effort. Currently, the most significant international collaboration is with India, although the collaboration is also forming ties with other European and Asian institutions. The earliest construction start dates for Project X is FY2016, with ongoing R&D now to refine the technical design and perform value engineering studies to optimize the performance/cost.

## Acknowledgments

This work is supported in part by the U.S. Department of Energy.

## References

[1] DOE vision for US HEP: http://www.er.doe.gov/hep/vision/index.shtml
[2] Fermilab Project X website: http://projectx.fnal.gov/
[3] Accelerators for America's future: http://www.er.doe.gov/hep/files/pdfs/Accel_for_Americas_Future_Final_Report.
econf/editors/eprint-template/instructions.html